# Surface engineering for cellulose as a boosted Layer-by-Layer assembly: excellent flame retardancy and improved durability with introduction of bio-based "molecular glue"

Can Fu [a,b,c], Xiaoli Xu [c], Guang-Zhong Yin [a], Baoyun Xu [c], Pingyang Li [c], Bo Ai [c], Zhongjie Zhai [d], Fei Gao [d], Jinguo Zhai [c], De-Yi Wang [a]*

[a] IMDEA Materials Institute, C/Eric Kandel, 2, 28906 Getafe, Madrid, Spain

[b] Universidad Politécnica de Madrid, E.T.S. de Ingenieros de Caminos, 28040 Madrid, Spain

[c] State Key Laboratory of Polyolefins and Catalysis, Shanghai Engineering Research Center of Functional FR Materials, Shanghai Research Institute of Chemical Industry Co. LTD., Shanghai, 200062, China

[d] Zhejiang Ruico New Material Co., LTD., Huzhou, 313018, China

**\*** Corresponding author

Email: deyi.wang@imdea.org

Tel: +34 917871888

**Abstract**

Layer-by-Layer (LbL) assembly was attractive as a versatile tool to address the flammability of cotton, while the washing fastness of LbL coating stayed an issue. Aiming to tackle this issue, LbL layers consisted of phenylphosphonic acid (PHA) and 3-aminopropyltriethoxysilane (APTES) was deposited on polydopamine (PDA)-coated cotton. The prepared cotton reached 31.4% of limiting oxygen index (LOI), and extinguished immediately after removing the ignitor. Peak of heat release rate (pHRR) attenuated around 36 % compared with pure cotton. A combined barrier and quenching mechanisms were proposed. Moreover, enhanced washing durability (24.1% of LOI) was achieved even after 50 detergent laundering cycles. A facile, boosted LbL approach with proposed π−π stacking interactions between PDA abundant aromatic structures and benzene ring in PHA from LbL layers, is first to put forward to construct durable efficient flame retardant (FR) cotton. This work attempted to enlighten more thoughts and design for durable FR cotton fabrics.

**Key words:** functional cotton, flame retardancy, durability, polydopamine, Py-GC-MS

.

# 1. Introduction

Given that cotton, being important natural fabrics, are prone to fire damages, both industrial and academic researchers have been making efforts to tackle this issue since last century [1-4]. Among the approaches to impart cotton FR property, Layer-by-Layer (LbL) assembly way was extraordinary attractive as a versatile tool. In this way, varieties of flame retardant (FR) compositions could be conferred to cotton via controlled construction by molecular. Nevertheless, LbL coating existed a general limitation of durability. This way was considered to an improvement of nanoparticles absorption on fabrics with feasibility in the early stage [5, 6]. Alongi et al. applied a series of LbL compositions onto cotton-based fabrics, achieving good flame retardancy with self-extinguish behavior [7-9]. Grunlan et al. fabricated varieties of polyelectrolytes and P-containing counterpart by LbL approach to render cotton FR and other desired functions [10, 11]. Nevertheless, above works didn't mention washing durability of LbL coating.

Generally, when it comes to LbL coating, most studies didn't address the washing ability. Only a few reports focused on improving the washing ability of LbL coating. For instance, Anna et al. synthesized a durable FR cotton fabrics which presented an improved washing fastness with release tests in static and dynamic conditions. The cross-linking action of $TiO_2$ phase strongly interacted both with DNA and cotton substrate [12, 13]. Hu et al. added hypophosphorous acid into polyethyleneimine layer and oxidized sodium alginate layer to achieve cross-linking structure between above polyelectrolyte layers [14]. Then the obtained fabrics kept good flame retardancy after

12 laundering processes with detergent. Washing fastness was imparted to fabrics by the cross-linking structures between polyelectrolyte layers in this work. Carosio et al. deposited 3 BL APP/chitosan layers on fabrics with assistance of UV-curable resin to obtain durable water-washed ability without detergent [15]. The resin cross-linked APP and chitosan between the layers after UV curing. Another example of simplified LbL coating with APP/ branchedpolyethyleneimine obtained LOI value of 20.0% after washing 6 h with detergent although the authors didn't explain the reason of durability [16].

Obviously, adhesive resin was not an ideal option because it would come with surely soared flammability as a detrimental effect for cotton. Crosslinking between layers was another main impulse to washing ability. However, chemical groups of phosphonate esters, amides, and esters normally functionalized as crosslinking agents, being likely to hydrolyze in an alkaline washing environment with detergent. Consequently, functional durable coating collapsed in 50 or even fewer laundering circles after treating with crosslinking agents, though [14, 17, 18]. What's more, the interaction between cotton substrate and LbL coating layers was quite crucial, which was constantly ignored and blurred in reports [7, 8, 19]. Plasma and corona treatments were applied to make substrate charged before electrolytes coating in some works [20, 21]. However, electrostatic attraction resulted from these treatments with intricate instruments demanding, was unstable and futile in the fierce laundering process, either. Undeniably, washing durability for functional fabrics continued being challenging. It is necessary to explore new approaches for durable coating on fabrics by use of hydrolytically stable interaction.

Polydopamine (PDA), mimicking the adhesive nature of mussels, is extensively applied to surfaces of almost all types of materials for photothermal therapy, bioimaging, drug delivery and so on. PDA coating exhibits excellent flexibility, controllability, and environmental benignity as so-called "molecular glue" in many literatures [22-27]. In detail, almost any solid-surface could be adhered robustly with no need of pretreatment which was taken charge of by amine and catechol groups in PDA structures, being precisely analogous to adhesive protein from mussels. PDA could be easily obtained in mildly alkaline environment from self-polymerization of dopamine. Notably, the abundant aromatic structures confer PDA excellently robust interaction with most of drugs for delivery. π−π stacking between PDA and benzene-ring structures from drugs was responsible for the strongly driving force in recent reports [22, 28, 29]. PDA-based materials were found sufficiently robust to attach drugs instead of departing in an alkaline environment more than 72 h [22, 30].

Since washing fastness of LbL coating stayed an issue for a long time, continuing endeavors such as adhesive resin assistance or crosslinking ether bond between LbL layers seemed to be mediocre improvements. Adhesive resin was not an ideal option because it would come with surely soared flammability as a detrimental effect. Crosslinking agents dominated by phosphonate esters, amides, and esters normally were likely to hydrolyze in an alkaline washing environment with detergent. Hydrolytically stable interaction between coating and substrate was needed. Additionally, PDA was widely implemented to be a robust carrier for drugs in an alkaline environment for a quite long time. In light of those consideration, we give a hypothesis that applying this "molecular glue" to work with aromatic components from LbL layers on the surface of cellulose can realize highly FR property with washing durability, simultaneously. In detail, PDA-coated cotton was easily obtained by adding pristine cotton into self-polymerization process of DA. Being a P-containing FR with aromatic structure, phenylphosphonic acid (PHA) was chosen as anionic layer, alternating with 3-aminopropyltriethoxysilane (APTES) as cationic layer to construct LbL system. Functional cotton with enhanced FR property and durability could be envisaged by above manipulations, marked as PDA-C-PHA-APTES. Herein, fire behaviors, washing fastness and corresponding mechanisms of this functional fabric will be investigated. This novel, environmental benign, sustainable, bio-based surface engineering as boosted LbL route that combines PDA-cover with alternate PHA/ APTES assembly layers, has never been proposed before. To the best of our knowledge, the concept of π−π stacking as a new approach to help improve durability for FR cotton with LbL coating was also first to put forward here.

## 2. Experimental

*2.1 Materials*

3-Aminopropyltriethoxysilane (APTES, ≥ 98%), phenylphosphonic acid (PHA, ≥ 98%), sodium hydroxide (NaOH, ≥ 98%), dopamine hydrochloride (DA, ≥ 98%), tris(hydroxymethyl)aminomethane (Tris, ≥ 99%) were all provided by Aladdin Chemical Co.. Cotton fabrics (104 g/m$^2$), was supported by Jinlin Fabric Co..

*2.2 Synthesis of PDA-coated cotton (PDA-cotton)*

Cotton fabrics were cleaned by deionized water and dried at 80 °C for 5 h. PDA-coated cotton was fabricated by a modified PDA self-polymerization process [31]. Dissolve 60 mg dopamine hydrochloride (DA) with 800 mL distilled water (0.0075 wt%). Then, add 0.968 g tris(hydroxymethyl)aminomethane (Tris) into above solution while stirring. After 10 min, pH was kept about 7.5 because of Tris. Later, pristine cotton was immersed in above solution for continuous stirring at room temperature for another 12 h. The treated cotton was squeezed to get rid of the excess solution and rinsed many times to ensure no more adhered solution. Subsequently, PDA-cotton was obtained after drying at 80 °C for 5 h.

*2.3 Preparation of functional fabrics*

PHA solution (3 wt%) was formed by dissolution PHA solid with deionized water. We then adjusted the pH~7 by adding 2M NaOH solution (PHA anionic solution). 5 wt% APTES solutions with pH = 5 was adjusted by adding 2 M nitric acid (APTES cationic solution) [32].

**PDA-C-PHA-APTES**. PDA-cotton was first immersed for 5 min in PHA anionic solution, and then dipped in APTES cationic solution for another 5 min, for forming 1 bilayer (BL). For the subsequently repeated dipping process, we only keep 1 min in both PHA anionic solution and APTES cationic solution cycle. Notably, in order to avoid the contamination of solution for next step and eliminate weakly absorbed chemicals of current step, the treated cotton was squeezed and rinsed three times respectively after each dipping step. After 5 bilayers coating, both the PHA and APTES solution would be renewed. The functional cotton samples were prepared after heating for another 2 h at 80 °C, until the desired bilayer numbers (1 BL, 2 BL, 8 BL) were obtained, which were marked as PDA-C-PHA-APTES 1 BL, PDA-C-PHA-APTES 2 BL and PDA-C-PHA-APTES 8 BL, respectively (**Fig. 1a**).

In order to investigate the washing durability of PDA-C-PHA-APTES 8 BL, reference samples PDA-C-APTES-PHA and C-PHA-APTES were prepared by following the similar preparation procedure. Notably, we adjust the coated bilayer number for each reference sample in order to keep the same weight loading of the coating as sample PDA-C-PHA-APTES 8 BL.

**PDA-C-APTES-PHA.** PDA-cotton was firstly immersed in APTES cationic solution for 5 min, and then dipped in PHA anionic solution for another 5 min, forming 1 BL coating. The subsequent steps followed the same procedure until 12 BL, which is marked as PDA-C-APTES-PHA (**Fig. 1b**).

**C-PHA-APTES.** First, pristine cotton was immersed in PHA anionic solution for 5 min, and then dipped in APTES cationic solution for another 5 min, forming 1 BL coating. In order to achieve the same weight loading, the subsequent steps followed the same procedure until 14 BL, which is marked as C-PHA-APTES **(Fig. 1c)**.

**180 PDA-C-PHA-APTES**. The higher amount of PDA coated cotton was obtained by using the higher DA concentration (0.02 wt%). Subsequently this higher PDA-coated cotton followed the same procedure as PDA-C-PHA-APTES aside from the number of bilayers, giving rise to sample 180 PDA-C-PHA-APTES (7 BL) with the same weight loading of PDA-C-PHA-APTES 8 BL.

**PDA-C-PHA.** PDA-cotton was immersed for 5 min in PHA anionic solution and then dried in oven for Continuous Wave Electron Paramagnetic Resonance test.

Different coating constitutes and ultimate weight loading were summarized in Table 1. The weight loading of treated cotton was calculated according to equation (1):

$$\text{Weight loading} = (W_{fc} - W_0)/W_{fc} \times 100\% \tag{1}$$

$W_0$ and $W_{fc}$ represented the weights of cotton fabrics pre- and post- functionalization treatment, respectively. The sample formulation details were showed in **Table S1**.

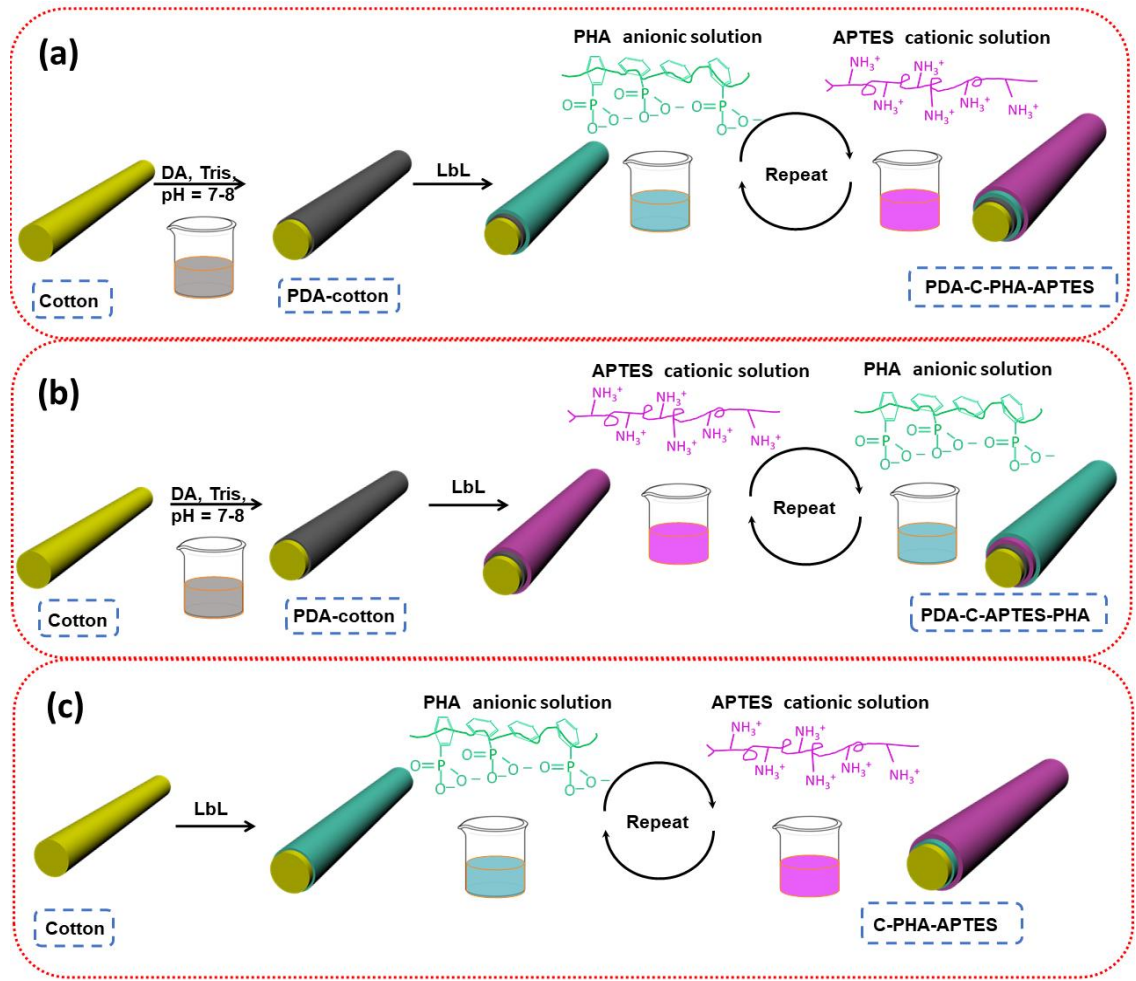

**Fig. 1.** The preparation process of different functional cotton samples.

*2.4 Measurements*

Scanning electron microscope (SEM) equipped with X-Ray microanalysis (EDX), Carl Zeiss. Fourier transform infrared spectroscopy (FTIR) was performed on a Nicolet iS50 spectrometer. NETZSCH TG209 F3 was used to perform thermogravimetric analysis (TGA) with a heating rate of 20 °C /min in air atmosphere from room temperature to 600 °C, with a flow speed of 20 mL/min. Najing Shangyuan analytical instruments Co., China, CFZ-4 was performed for vertical flame test (VFT). Specimen sheets for testing were of 300 × 80 mm$^2$ (GB/T 5455-2014). In this test, 5 specimens were tested under a methane flame for 12 s each time. Limiting oxygen index (LOI) was tested with dimensions of 150 × 58 mm$^2$ according to a standard GB/T 5454-1997 (Fire Testing Technology Co., UK). FTIR spectrometer connected with a thermogravimetric analyzer (PerkinElmer Co., USA.) was used for the thermogravimetry-Fourier transform infrared spectrometry (TG-FTIR). The productes resulted from heating with 10 °C /min rate in nitrogen to 600 °C, were directed to FTIR sample chamber by an insulating pipe. Thermo ESCALAB 250Xi spectrometer fitted with an X-ray source (Al Kα with photon energies 1486.6 eV) was used to have X-ray photoelectron spectroscopy (XPS) analysis. FTT Cone Calorimeter instrument was conducted to test fire hazards of samples according to ISO5660-1 standard under a 25 kW/m$^2$ heat flux with size of 100 × 100 × 1.6 mm$^3$ (5-layer fabrics). A wire grid was used to prevent expanding samples in the test. Thermo Fisher Scientific Co., USA was used for Raman spectra with a DXR micro-spectroscopy system. Thermo TRACE 1310 for GC and ISQ 7000 for MS with Frontier EGA/PY-3030D type of pyrolyzer were performed to have pyrolysis gas chromatography mass spectrometry (Py-GC-MS) test. Products produced at 600 °C under He atmosphere. Washing durability was tested based on AATCC 143-2014 method. Hand wash was applied to this work according to the standard washing process. 30.3 ± 0.1 g of 2003 AATCC Standard Reference Liquid Detergent was used to 7.57 ± 0.06 L water at 41 ± 3 °C. Three pieces of cotton sample were added into the above washing solution for 2 min, and then rinse samples with 7.57 ± 0.06 L water at 41 ± 3 °C. The whole above process was defined as one circle. The above cotton samples were washed for 1, 5, 30, 50 times, respectively. The weight loading was calculated and LOI was tested after rinsing and drying. Washing water with detergent from 4 different samples after 1 circle and 5 circles was collected to do Inductively Coupled Plasma Mass Spectrometry (ICP-MS) test. Nexion 300D inductively coupled plasma mass spectrometer was used to do ICP-MS test with method SCAI_EN0050. The Continuous Wave Electron Paramagnetic Resonance (CW-EPR) measurement were performed with Bruxer A300. The modulation amplitude was 1 G, the microwave power was 11.79 mW, and the microwave frequency was measured 9 GHz. Three comparative samples were prepared for CW-EPR test: pristine cotton fabrics, PDA-cotton, and PDA-C-PHA. Based on GB/T 3923.1-2013, the tensile strength and breaking elongation of cotton samples was evaluated with Instron Tension tester. In this test, five replicates with 25 × 5 $cm^2$ were prepared. The drawing speed was 20 mm/min at a clamping distance of 20 cm.

## 3. Results and Discussion

*3.1 SEM morphology and chemical composition of functional cottons*

SEM was carried out to distinguish surface morphologies of prepared cotton samples. In **Fig. 2 (a)**, the surface of cotton displayed a little rough. When it comes to PDA-cotton, the morphology exhibited completely uniform and conformally homogeneous. This PDA-coating was not easy to distinguish from the pristine cotton aside from the accidently broken part of coating under microscope in **Fig. 2 (b)**, since the coating layer maintained adequately thin and smooth. After 8 BL of PHA/APTES deposition, a continuous coverage on cellulose could be seen for PDA-C-PHA-APTES

8 BL. Because of the increased coating layers, some aggregates of PHA/APTES appeared on the surface, seen in **Fig. 2 (c)**. As the EDX mapping showed in **Fig. 2 (d) (e)**, the chemical compositions of PDA-cotton (C, O, N) differed from PDA-C-PHA-APTES 8 BL sample (P, Si, C, O, and N), although N element was almost invisible under SEM, generally [33]. Meanwhile, similar results were well achieved in XPS spectra in **Fig. 3 (b)**. Furthermore, a related quantitative analysis of LbL coating was provided in **Table S2**.

To further demonstrate the indeed deposition of different coatings on cotton, FT-IR and high-resolution XPS spectra of N 1s were carried out. Those typical functional groups of samples make the spectra of Cotton, PDA-cotton and PDA-C-PHA-APTES 8 BL samples differ in **Fig. 3 (a)**. In detail, a new peak at 554 $cm^{-1}$ (N-H out-of-plane bending) in spectra of PDA-cotton appeared, not being displayed in cotton. Analogously to what observed for FT-IR curves, the peak around 399.7 eV (N 1s) for PDA-cotton in **Fig. 3 (b)** could be obtained compared with cotton, indicating successful PDA layer formed on the surface of cotton [34, 35]. Moreover, for PDA-C-PHA-APTES 8 BL, the absorption peaks at 1537 $cm^{-1}$ and 1233 $cm^{-1}$ were attributed to the stretching vibration of N-H [36, 37] and P=O, which resulted from APTES and PHA, respectively [38, 39]. Another characteristic absorption peak could be at 964 $cm^{-1}$ (Si–O bending) [40, 41]. In particular, stretching vibrations of Si-C located around at 782 $cm^{-1}$ [42, 43]. The absorption peaks at 690 $cm^{-1}$ could be bending vibration mode of C–H for aromatic ring from PHA [35]. Notably, both PDA-cotton and PDA-C-PHA-APTES 8 BL exhibited N-containing group in XPS and FT-IR spectra compared with pristine cotton, so the high-resolution N 1s spectrum was checked in **Fig. 3 (c)**. Obviously, pristine cotton didn't show any absorption during this characteristic area. For PDA-cotton, the peak at 399.7 eV was solely dominated by indole group on account of self-polymerization process of dopamine [22, 44, 45]. By contrast, the spectra of N 1s for PDA-C-PHA-APTES 8 BL, could be deconvoluted to two peaks, 399.6 eV (ascribed to PDA) and 401.4 eV (typically protonated amine due to APTES) [22, 46]. Therefore, combined SEM, FTIR and XPS results, PDA conformally coated pristine cotton in first step, and PHA/APTES assembly layers can be further deposited on PDA-cotton successfully.

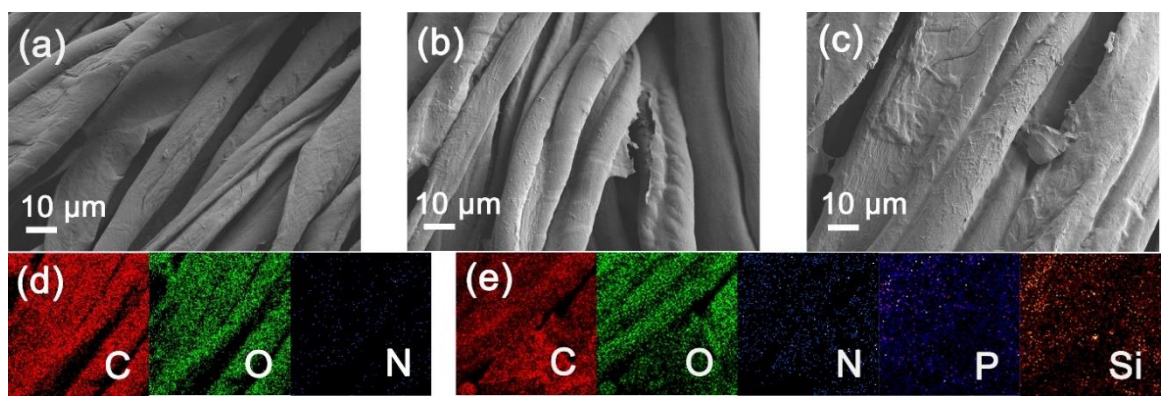

**Fig. 2.** SEM images of (a) cotton, (b) PDA-cotton, (c) PDA-C-PHA-APTES 8 BL and corresponding EDX mapping of (d) C, O and N elements of PDA-cotton, (e) C, O, N, P and Si elements for PDA-C-PHA-APTES 8 BL.

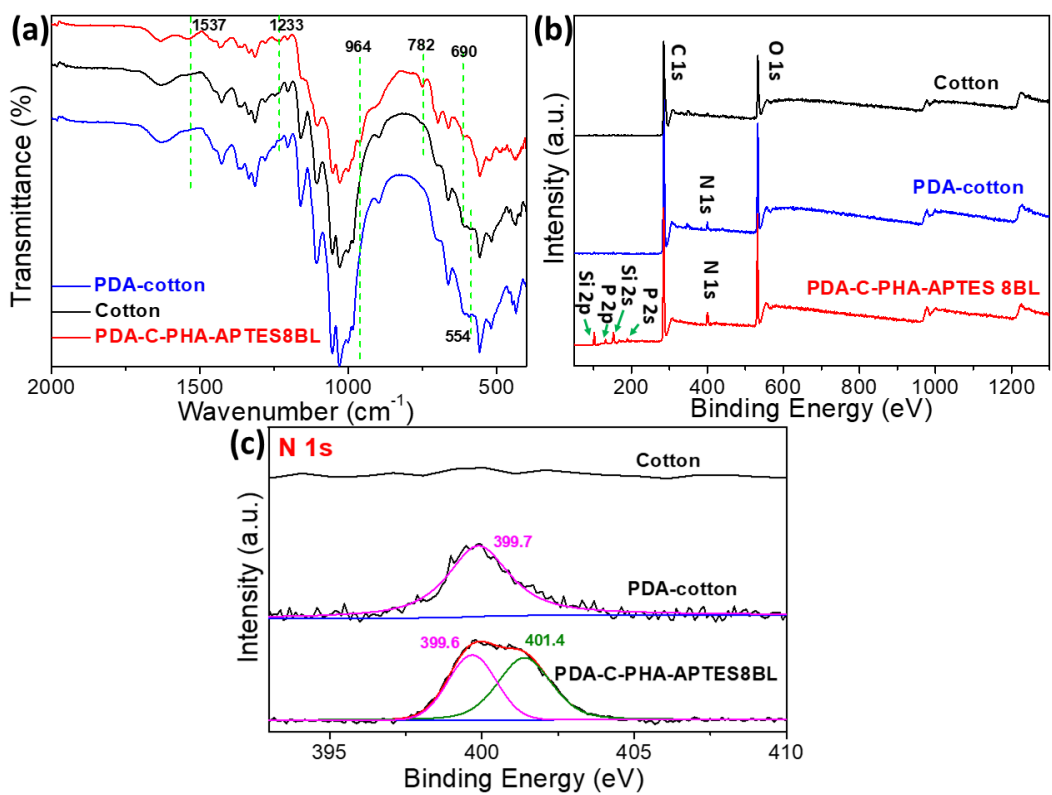

**Fig. 3.** (a) FT-IR, (b) XPS spectra and (c) high-resolution N 1s spectrum for Cotton, PDA-cotton, and PDA-C-PHA-APTES 8 BL samples.

*3.2 TGA of cotton samples*

The thermal-oxidative decomposition of cotton was considered very complicated, shown in **Fig. S1** and **Table S3.** Cotton fabrics were documented to display decomposition process in three steps: (1) the generation of volatile products and aliphatic char ranging from 300 to 400 °C, (2) conversion from aliphatic char to aromatic ones resulting in carbonization and simultaneous oxidation of residue to yield CO or $CO_2$ above 400 °C, (3) further oxidation for hydrocarbon species to $CO_2$ mostly at 800 °C [47-49]. Based on this, two decomposition peaks were found around at 351 and 482 °C for all cotton samples except PDA-C-PHA-APTES 8BL in DTG curves in

**Fig. S1 (b)**, respectively. Indeed, for PDA-C-PHA-APTES 8BL curve, the second step decomposition completely disappeared, which was ascribed to a higher carbonization effect in the concurrent presence of the silicon and phosphorous species [40, 48]. This phenomenon was caused by directly catalyzing to aromatic char and thermally stable Si, P containing ceramic residue instead of experiencing the second step of the conversion of aliphatic char. Notably, $T_{max1}$ increased slightly, which is in good agreement with the report that PDA was considered to improve the stability of FR system in high-temperature area [50]. The degradation rate (71.63 wt%/min) of PDA-cotton was even higher than that of cotton. As exhibited in **Fig. S1 (a)**, no residue left for both cotton and PDA-cotton samples at 600 °C. With the increasing LbL layers, the residues soared a lot (8.5, 13.0, and 40.3% for PDA-C-PHA-APTES 1, 2, and 8 BL, respectively). TGA results indicated PDA-C-PHA-APTES samples performed improved thermal stability and carbonization ability.

*3.3 Fire test*

Fire behaviors of all the cotton samples were investigated by LOI test and VFT visually. Digital photos were recorded in **Fig. 4**, and corresponding data were tabulated in **Table S4**. From the screen shot after ignition, both pristine cotton and PDA-cotton burned completely in impinging flame with no residue, seen in **Fig. 4 ($a_1$) ($a_2$), ($b_1$) ($b_2$)**. LOI values of cotton and PDA-cotton were 17.4% and 17.7%, respectively, which indicated PDA-coating scarcely influenced fire behavior of cotton. While, in **Fig. 4 ($e_1$) ($e_2$)**, the flame extinguished at once with no after-flame for PDA-C-PHA-APTES 8 BL after removing the ignitor. The damage length was solely 9.5 cm with maintaining fabric integrity. Both PDA-C-PHA-APTES 1 BL and PDA-C-PHA-APTES 2 BL displayed after-glow suppression and char-forming capacities although continued burning after flame, seen in **Fig. 4 ($c_1$) ($c_2$), ($d_1$) ($d_2$)**. In addition, the value of LOI incremented along the increased number of bilayers of PHA/APTES, seen in **Table S4**. In particular, LOI value of PDA-C-PHA-APTES 8 BL remarkably soared to 31.4%.

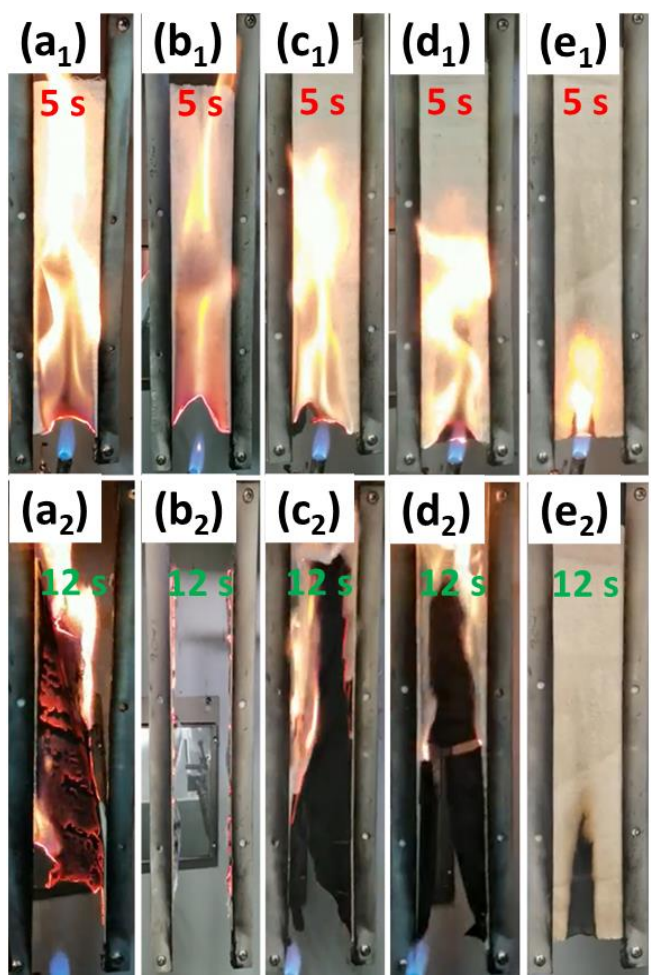

**Fig. 4.** Digital photos of cotton samples in vertical flame test after 5 s and 12 s: ($a_1$, $a_2$) Cotton, ($b_1$, $b_2$) PDA-cotton, ($c_1$, $c_2$) PDA-C-PHA-APTES 1 BL, ($d_1$, $d_2$) PDA-C-PHA-APTES 2 BL, ($e_1$, $e_2$) PDA-C-PHA-APTES 8 BL.

To fully discuss the fire safety of the PDA-C-PHA-APTES 8 BL in fire scenario,

the cone calorimeter test (CCT) was used, seen in **Fig. 5** and **Table S5**. The peak value of HRR (pHRR) for PDA-C-PHA-APTES 8 BL reduced by 36% in contrast with that of cotton. Value of total heat release (THR) for PDA-C-PHA-APTES 8 BL decreased by 56% from 10.6 MJ/m$^2$ of cotton to 4.7 MJ/m$^2$. Furthermore, the value of time to ignition (TTI) of PDA-C-PHA-APTES 8 BL significantly delayed from 50 s (Cotton) to 70 s, demonstrating PDA-C-PHA-APTES 8 BL was more difficult to be ignited. This phenomenon caused by synergistic effect of silicon and phosphorus in FR system, as also documented by other reports [48, 51-53]. The weight loss rate of PDA-C-PHA-APTES 8 BL has been declining along the time in comparison with cotton, seen in **Fig. 5 (b)**. In **Table S5**, the char residue of PDA-C-PHA-APTES 8 BL (39.1%) remarkably increased compared with 2.8% of cotton. This observation was well depicted in TGA analysis, indicating the formation of consistent and coherent residues after testing. The pristine cotton was almost consumed after CCT while PDA-C-PHA-APTES 8 BL maintained their shapes and cellulose structures in left and right image of **Fig. 5 (d)**, respectively. By contrast, a sharp increment for total smoke production (TSP) value with presence of the PHA/APTES coating, which was attributed to the incomplete combustion of FR samples [54, 55]. However, the decline of fire growth rate index (FIGRA referring to dividing PHRR by time to PHRR) from 1.5 to 0.9 kW/m$^2$ s displayed that the fire developing was prohibited. The value of average effective heat of combustion (Av-EHC, referring to burning rate of volatiles in gas-phase) of PDA-C-PHA-APTES 8 BL slightly decreased to 12.9 MJ/kg referring to 15.1 MJ/kg of pristine cotton, implying possible gas-phase FR mechanism. All above VFT, LOI and CCT results demonstrated functional fabric PDA-C-PHA-APTES 8 BL exhibited FR property in an efficient way.

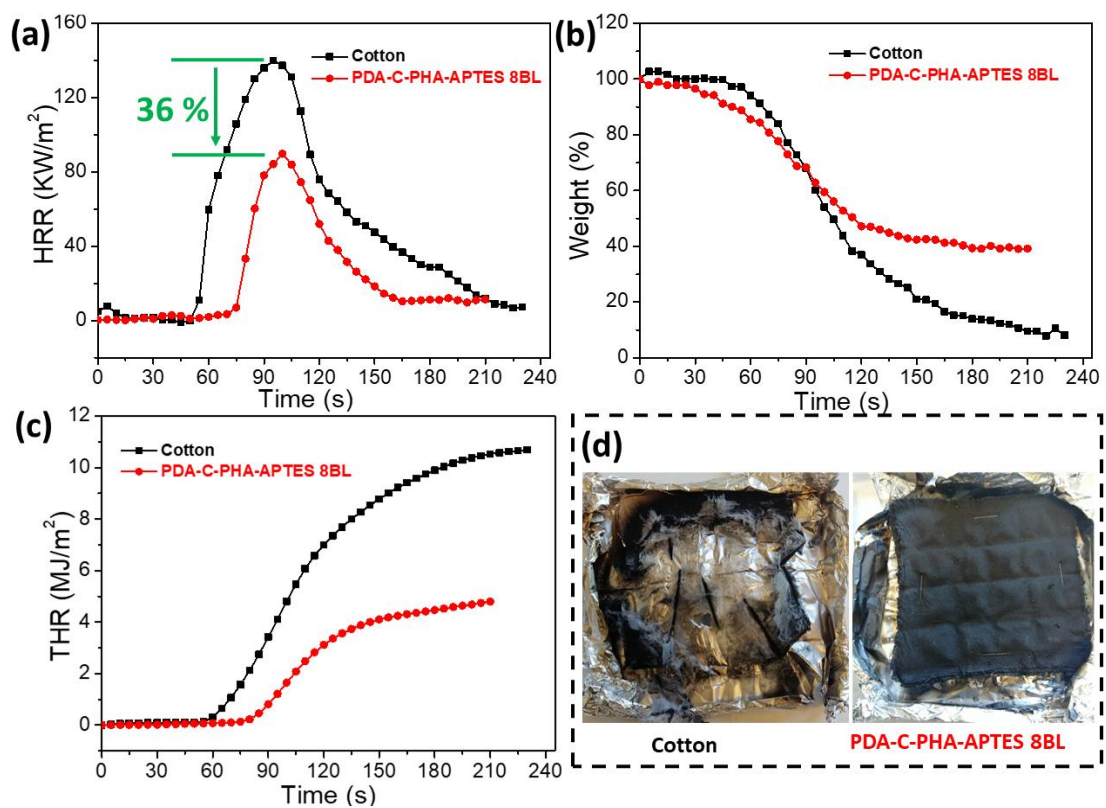

**Fig. 5.** Summary of results obtained from cone calorimeter test: (a) HRR curves, (b) weight loss curves, (c) THR curves of Cotton and PDA-C-PHA-APTES 8 BL; (d) digital photos of residues after CCT: cotton (left), PDA-C-PHA-APTES 8 BL (right).

*3.4 Char Residue Study*

FR property of cotton was strongly influenced by the composition and quality of char after combustion. So, SEM, Raman, FTIR and XPS were applied to do char analysis of cotton and PDA-C-PHA-APTES 8 BL. The residue of cotton was quite fragile after burning, seen in **Fig. 6 (a$_1$)** and **(a$_2$)**. However, the residue of PDA-C-PHA-APTES 8 BL was much more continuous and compact than that of cotton, which maintained the cotton structure and cellulose integrity. Notably, as showed in **Fig. 6 (b$_1$)** and **(b$_2$)**, the formation of obvious bubbles, being trapped in the swelling residue, proved the presence of intumescent char layer. In this FR system for PDA-C-PHA-APTES 8 BL, PHA worked as acid source, generating polyphosphoric acid during the combustion, which dehydrated cotton and promoted carbonization. Blowing agent was mostly considered to $NH_3$ from APTES as diluting effect. The cellulose units could serve as carbon source. The similar consequences were exhibited in Raman spectra. The value of $I_G/I_D$ for PDA-C-PHA-APTES 8 BL was 0.66, which increased in contrast with that of cotton (0.29), seen in **Fig. 6 (c) (d)**. That meant the char of PDA-C-PHA-APTES 8 BL had a higher graphitization degree, as G and D bands were respectively assigned to the vibrations of graphitic carbon and disordered carbon. Both SEM and Raman results could demonstrate the char of FR cotton PDA-C-PHA-APTES 8 BL displayed a sufficient barrier effect to isolate heat and oxygen, impeding flame spread.

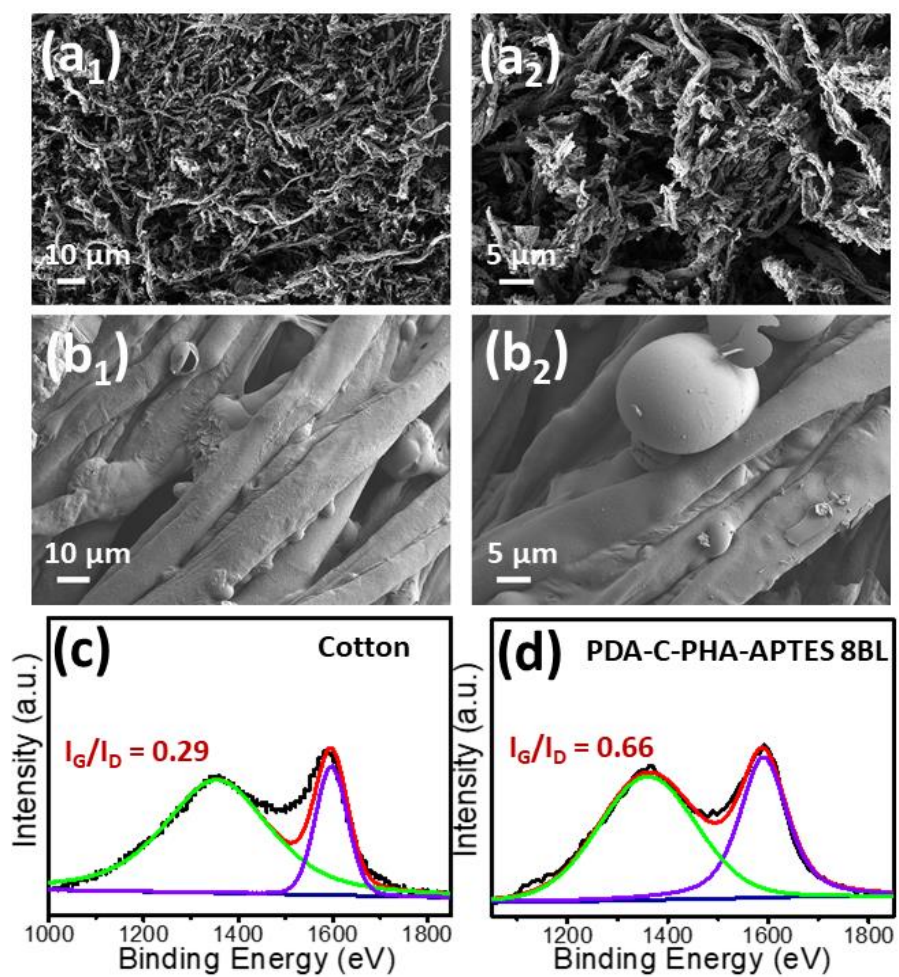

**Fig. 6**. SEM images of the char residues of different cotton samples after the vertical flame test: ($a_1$, $a_2$) cotton, ($b_1$, $b_2$) PDA-C-PHA-APTES 8 BL, Raman spectra of (c) cotton and (d) PDA-C-PHA-APTES 8 BL.

FTIR and XPS measurements were carried out to analyze composition of char. In **Fig. S2 (a)**, absorption bands around 1692, 1610 and 1435 cm$^{-1}$ were determined to the basic skeleton vibration of benzene ring. Besides, characteristic absorption peaks at 746, 709, 698 cm$^{-1}$ could be C–H bending vibration in benzene ring. All of these revealed the presence of the aromatic carbonaceous structure [35, 56]. Peaks at 1350 and 1105 cm$^{-1}$ referred to P=O and P-O-P/P-O-C, respectively [35, 57]. Correspondingly, in **Fig.**

**S2 (b)**, P 2p spectra was deconvoluted into P-O-C/P-O-P (134.6 eV) and O-P=O (133.8 eV) [58]. Another characteristic absorption peak around 921 cm$^{-1}$ was indicated to Si–O vibration [40]. Analogously to what observed from FTIR, thermally stable Si, P containing ceramic char was further demonstrated according to deconvoluted Si 2p spectra at 104.0 eV (Si-O), 103.4 eV (Si–P) and 101.7 eV (Si–C) [36, 59]. Moreover, peak at 540 cm$^{-1}$ being considered to N-H (out-of-plane bending) [60] indicated that nitrogen was also partly anchored in the char residues. Correspondingly, the spectra of N 1s could be deconvoluted to N–C (398.8 eV), N=C (400.5 eV), and quaternary/oxidized nitrogen (401.8 eV) [36]. Therefore, according above analysis, the continuous and compact char of PDA-C-PHA-APTES 8 BL was an intumescent layer composed of P/Si/N-containing ceramic structures and aromatic species which confined heat and oxygen transferring.

*3.5 Pyrolysis volatiles analysis*

With the aim of further investigating FR mechanism of this functional FR cotton, we did the pyrolysis volatiles analysis of PDA-C-PHA-APTES 8 BL which was characterized via Py-GC-MS and TGA-FTIR. As seen in **Fig. 7 (a)**, we focused on main pyrolysis products from cotton located at 1.53, 10.13, and 14.85 min although more than 40 peaks were detected. Based on the NIST data library, peak area around 48.49% at 14.85 min was ascribed to levoglucosan which was the main dehydration product from cotton. In fact, decomposition of levoglucosan was considered to general formation of highly flammable volatiles as reported in many literatures [47, 61]. Peaks at 1.53 min indicated the generation of $CO_2$ and $H_2O$ during the dehydration. Alcohol, aldehyde, and ketone released at 10.13 min. According to the GC and MS data, we could see main pyrolysis products of cotton were combustible volatiles. By contrast, PDA-C-PHA-APTES 8 BL decomposed in a completely different way with several new peaks detected in **Fig. 7 (b)**. At 14.79 min, the generation of levoglucosan decreased to 18.28% compared with 48.49% of cotton. Correspondingly, the release of water decremented at 1.53 min where main detectable volatiles were $NH_3$ and $CO_2$, seen in the MS spectra (m/z = 17 and 44). In Horrocks' theory and related researches [47, 61-64], carbonization and char-forming process played the most important role as condensed phase active FR. Therefore, furan, furfural, and 1, 4 :3, 6-dianhydro-α-D-glucopyranose, being main products at 12.23 and 10.56 min in **Fig. 7 (b) (c)**, remarkably increased as char intermediates instead of levoglucosan generation. Formation of furan ring and cyclodehydration from intramolecular carbohydrates dehydration happened under heated acidic condition. Moreover, peaks at 6.93 and 17.25 min were determined to be N and Si containing products, in analogy with char study of XPS and FTIR. Notably, peaks at 2.29 and 5.59 min were identified as P-containing volatiles according to NIST data. In particular, the m/z value of 63 and 94 appeared in MS spectra in **Fig. 7 (c)**, as the main products at 2.29, 2.44 and 3.43 min, were attributed to ·$PO_2$ and ·$P_2O_2$ free radicals respectively from the ·PO free radicals [65-68]. Furthermore, the presences of less reactive aromatic structures were demonstrated at 2.44, 3.43, 5.59, and 12.57min. Similar results were showed in **Fig. S3** from TG-FTIR results. The flammable pyrolysis volatiles of PDA-C-PHA-APTES 8 BL, such as hydrocarbons (2936 cm$^{-1}$, shown in **Fig. S3a**), aliphatic ethers (1084 cm$^{-1}$, **Fig. S3b**), CO (2184 cm$^{-1}$, **Fig. S3c**), and carbonyl compounds (1750 cm$^{-1}$, **Fig. S3d**) were greatly reduced in contrast with cotton [35, 69]. Alongside this, less flammable pyrolysis products, aromatic compounds, notably increased compared with cotton [69].

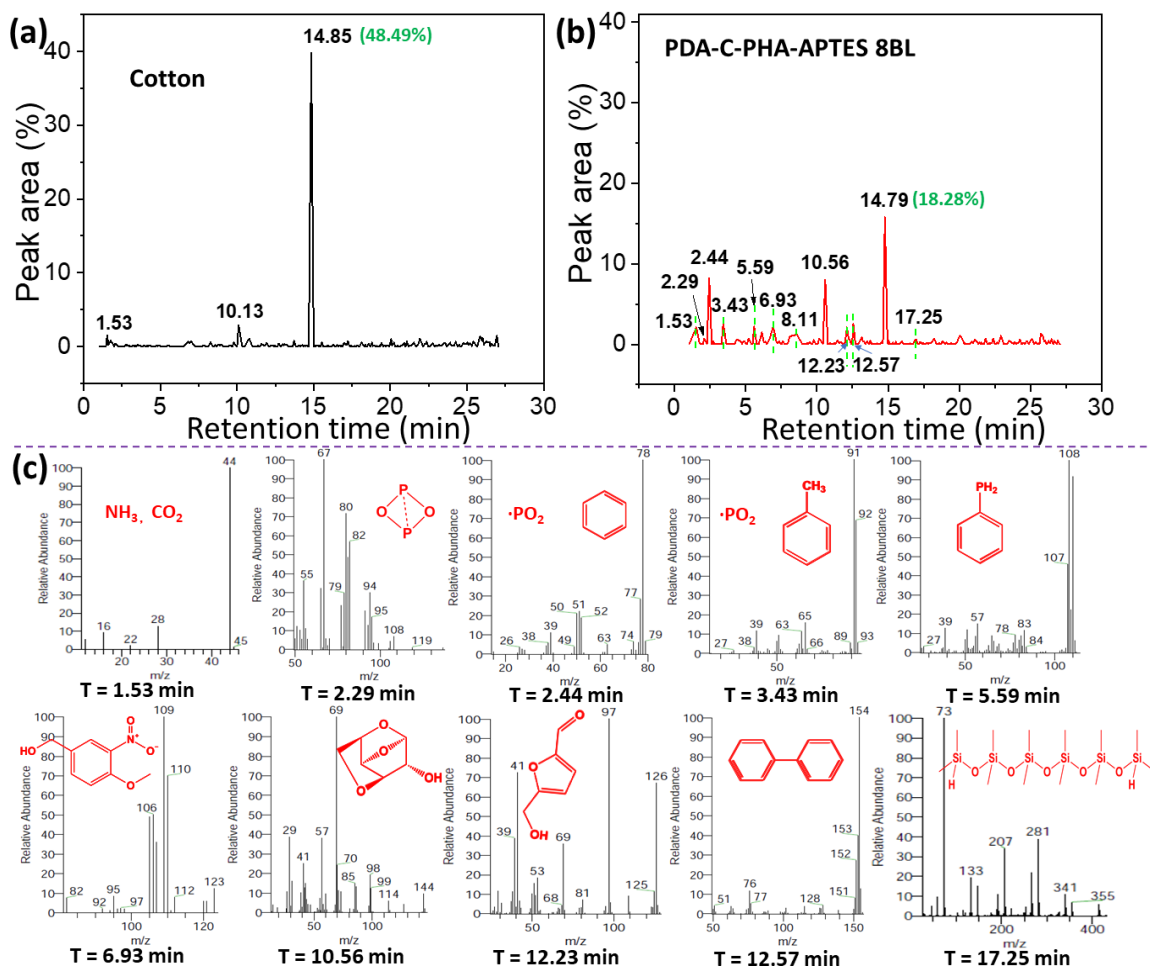

**Fig. 7**. GC spectra of (a) Cotton, (b) PDA-C-PHA-APTES 8 BL, (c) MS spectra of PDA-C-PHA-APTES 8 BL at different retention time.

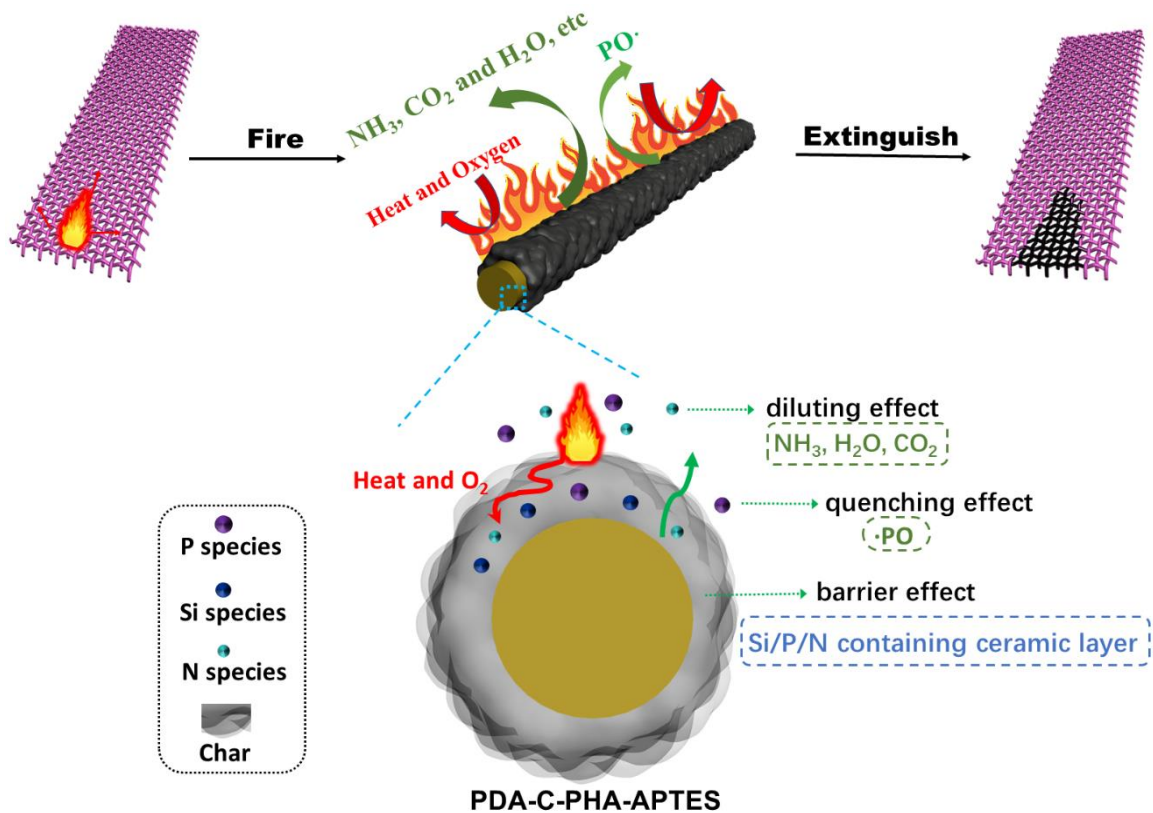

**Fig. 8**. Schematic illustration of proposed FR mechanism of functional cotton PDA-C-PHA-APTES 8 BL.

The TGA, VFT and LOI results illustrate PDA coating could be ignored in FR process since it scarcely influenced fire behavior of cotton at such a low content. So, based on all the analysis, a FR mechanism for this functional cotton PDA-C-PHA-APTES 8 BL combined gas phase and condensed phase mechanism were proposed in **Fig. 8**. PDA-C-PHA-APTES 8 BL released low-energy ·PO free radicals in the gaseous phase during combustion. These ·PO free radicals could capture high energy radicals such as ·H or ·OH radicals from the decomposing cotton, resulting in inhibition of the flame. Meanwhile, $H_2O$, $CO_2$, $NH_3$ and some nonflammable volatiles were produced to dilute the flammable volatiles and oxygen, and also cool down the temperature during the fire [43]. In the condensed phase, as the temperature rose, Si and P from LbL coating promoted the dehydration and carbonization process significantly [40, 48]. Simultaneously, an intumescent layer composed of P/Si/N-containing thermostable ceramic structures and aromatic species formed as a good physical barrier, which effectively hindered the penetration of oxygen and transferring of heat [40, 48]. As a consequence, this obtained FR cotton fabrics exhibited efficient FR property.

*3.6 Washing durability*

Washing ability of all the cotton samples was explored with the assistance of LOI test after washing with ionic detergent (2003 AATCC Standard Reference Liquid Detergent). In order to investigate the washing durability mechanism, we prepared some comparative samples PDA-C-APTES-PHA, 180 PDA-C-PHA-APTES, and C-PHA-APTES, shown in section 2.3. In the following part, PDA-C-PHA-APTES was used to represent PDA-C-PHA-APTES 8 BL for being concise. The amounts of PHA and APTES in washing water with detergent after different circles were evaluated directly from P and Si concentrations, respectively. According to **Fig. 9** and **Table S6,** most of coating was lost in the first five washing cycles, which was also verified by recent reports [12, 13]. So, washing water with detergent from above different samples after 1 cycle and 5 cycles was collected to do ICP-MS test. The results were shown and summarized in **Fig. S4** and **Table S7**.

In **Fig. 9** and **Table S6**, PDA-C-PHA-APTES exhibited good durability with maintained LOI value as 20.9%, being still superior to pristine cotton even after 50 cycles. Then weight remained on this cotton was 7.5 %. In fact, most of coating was lost in the first five washing cycles. Accordingly, the washing water from PDA-C-PHA-APTES exhibited low release of P (52.8 ppm) and Si (96.3 ppm) after 1 washing cycle in **Fig. S4**. The release of P slightly increased from 52.8 ppm to 68.7 ppm after 5 washing cycles, but nearly 2-fold increment of Si release (from 96.3 to 175.3 ppm) was obtained. That phenomenon could be explained that the interaction inside LbL layers was electrostatic attraction, which was peeled off by the detergent washing, as shown in **Fig. 10 (d)**. However, the primarily several bi-layers were stacked on cotton robustly because of π-π stacking between aromatic structures in PHA and benzene rings in PDA [22, 23, 29, 30]. As a result, the Si release in the washing water after 5 cycles was more than P release. By contrast, PHA/APTES bilayers were deposited on cotton directly without PDA (C-PHA-APTES) displayed no durability at all after 5 washing cycles. Moreover, C-PHA-APTES exhibited 17.9% of LOI value and 1.1% of weight loading even solely after 1 cycle. Meanwhile, the washing water of C-PHA-APTES exhibited highest loss of P and Si. The interaction for C-PHA-APTES between PHA/APTES bilayers and cotton substrate was considered common physical adsorption, shown in **Fig. 10 (f)**. However, the weight loading and LOI value of 180 PDA-C-PHA-APTES soared to 9.3% and 24.1% respectively after 50 washing procedures with increasing PDA amount. And the release of both P and Si for washing water of 180 PDA-C-PHA-APTES showed the lowest value after 1 and 5 washing cycles. These results suggested that more π-electron-rich surfaces which was supported by increasing PDA-coating could intensely load more LbL coating after washing. Notably, for PDA-C-APTES-

PHA, the release of P and Si in washing water displayed less than that of C-PHA-APTES, a certain extent of durability with 18.4% of LOI exhibited after 50 cycles. This rare durability demonstrated that APTES layers only physically absorbed on the PDA coating. And still this slight durability may come from the interaction between few following PHA layers and PDA coating which was not completely covered by APTES layers, shown in **Fig. 10 (e)**.

In order to prove the direct interaction between PDA-coating and PHA, Continuous Wave Electron Paramagnetic Resonance (CW-EPR) measurement was employed. Most commonly, X-band (9 GHz) EPR has been used as a "fingerprint" technique, revealing characteristic radicals. In **Fig. S5**, the curves of pristine cotton didn't show any signal under magnetic field. By contrast, PDA-cotton show a slightly asymmetrical signal with g value of 2.004, which is in accordance with reports [70, 71]. Two radicals were demonstrated to coexist in this system. One radical comes from semiquinone structure in PDA. The second radical is located in the benzene ring of PDA. The dominant radical that contributes to that signal is reported to aromatic ring π-radical [70]. After incorporating PHA with PDA cover (PDA-C-PHA), the signal disappeared in **Fig. S5**. So, EPR test verified the direct interaction between PDA-cover and PHA layer. Based on the reports, the main interaction in this system was π-π stacking between benzene ring of PDA and aromatic structures in PHA.

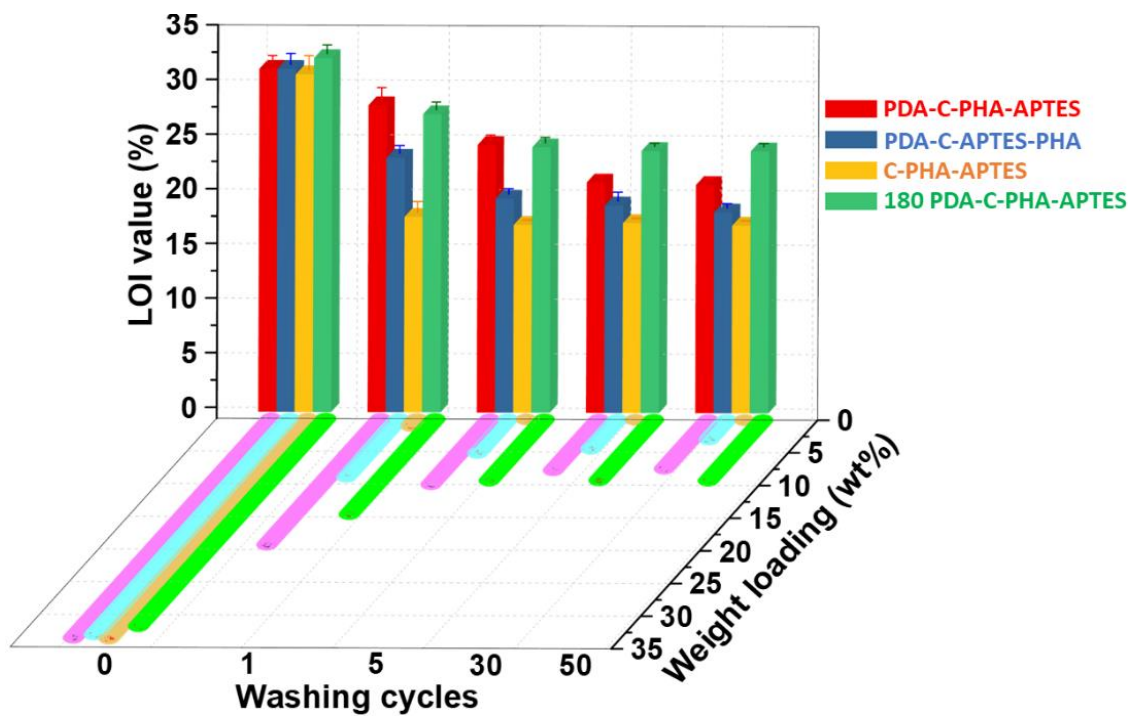

**Fig. 9**. Weight loadings and LOI values of PDA-C-PHA-APTES, PDA-C-APTES-PHA, C-PHA-APTES, 180 PDA-C-PHA-APTES after washing 1, 5, 30, and 50 cycles.

In conclusion, for functional cotton PDA-C-PHA-APTES in this work, PDA coated cotton by self-polymerization of DA on the surface of cotton, seen in **Fig. 10 (a) (b)**. Then PHA was intensively attached to abundant benzene rings on PDA-cotton via π-π stacking in **Fig. 10 (c)**. APTES was packed to PHA by electrostatic interaction subsequently in **Fig. 10 (d)**. As is known, 50 washing cycles with ionic detergent were a severe washing test for functional cotton. Therefore, above washing test demonstrated π-π stacking between PDA-coated cotton and aromatic structures in LbL coating can impart FR cotton washing fastness with 50 cycles.

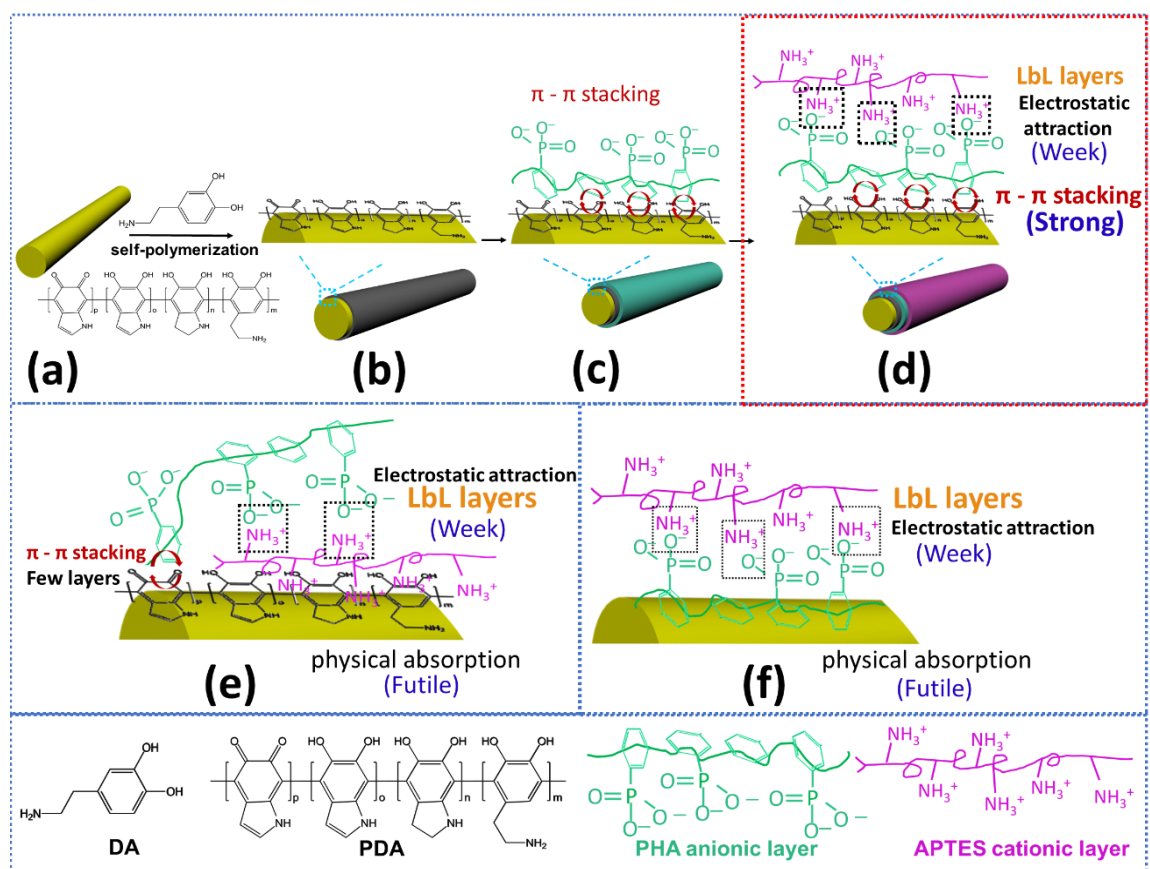

**Fig. 10**. Scheme of fabrication for cotton samples, (a) pristine cotton, (b) PDA-cotton, (c) PHA anionic layer attached on PDA-cotton, (d) durable FR cotton PDA-C-PHA-APTES, (e) PDA-C-APTES-PHA, (f) C-PHA-APTES.

*3.7 Tensile strength*

Mechanic property of cotton samples are obtained in **Fig. S6.** After PDA coating, both tensile strength (10% in warp and 19% in weft higher) and breaking elongation (1% in warp and 6% in weft higher) increased in contrast to cotton. That observation may result from the conformal coating by PDA as a protective layer for glycoside bonds of cellulose. Moreover, tensile strength of PDA-C-PHA-APTES in warp and weft directions had slight increases by 3% and 15% respectively compared with that of cotton. However, for breaking elongation, PDA-C-PHA-APTES had sight decrease in weft direction by 11%. This small reduction of breaking elongation might be resulted from introduction of acid in PHA with probably breaking a few glycoside bonds of cotton cellulose [41, 72].

## 4. Conclusion

In this work, functional cotton with excellent FR property and improved washing ability was achieved by depositing alternate PHA/APTES bilayers on PDA coated cotton. The prepared cotton reached as high as 31.4% for LOI value, and extinguished immediately after removing the ignitor in vertical fire test. Value of the pHRR attenuated around 36 % compared with pristine cotton. A combined barrier effect and quenching effect mechanism were proposed for this FR system. The gas phase flame retardancy was realized via free radical quenching effect and diluting effect by nonflammable volatiles ($NH_3$, $CO_2$, $H_2O$). Moreover, an intumescent layer consisted of aromatic carbonaceous structures and P/Si/N thermostable ceramic char layer provides good barriers to protect the underlying cotton, which effectively hindered penetration of oxygen and transferring of heat in condensed phase. Hence, obtained cotton exhibited excellent FR property eventually. In addition, FR property of PDA-C-PHA-APTES was still superior to pristine cotton after 50 detergent washing process. In particular, 180 PDA-C-PHA-APTES maintained 24.1% of LOI after 50 detergent washing cycles. The robust LbL coating on cotton relied on π-π stacking between aromatic structures in PHA and benzene rings in PDA. Overall, a novel, environmental benign, sustainable, bio-based surface engineering as boosted LbL route was provided here. The hypothesis with construction of excellent FR property and improved durability of cotton by surface engineering as a boosted LbL assembly was realized indeed. Although further improvement for washing fastness by enhancing the interaction inside LbL layers would be considered in the future, still this work attempted to enlighten new thoughts and design to fabricate efficient and durable FR cotton. The concept of π−π stacking as a new, promising approach may be applied by more versatile FRs containing aromatic structures and other kinds of fabric substrate like PET or nylon with bio-based PDA coating.

## Acknowledgement

This research is partly supported by Natural Science Foundation of Science and Technology Commission of Shanghai Municipality (19ZR1423700), Shanghai Engineering Research Center of Functional FR Materials (19DZ2253700), and China Scholarship Council (201506950020).